\documentclass[aps,prb,twocolumn]{revtex4}
\usepackage{graphicx}

\begin{document}
\title{Thickness Dependence of the Reorientation Phase Transition}
\author{David Clarke}
\affiliation{Johns Hopkins University, Department of Physics and
Astronomy, 3400 N. Charles St., Baltimore, Maryland 21218}

\begin{abstract}
    This report examines the thickness dependence of the
    reorientation phase transition in ferromagnetic films with
    perpendicular anisotropy. That is, we find the exact boundary of
    metastability of uniformly magnetized in-plane states as the
    solution to a set of transcendental algebraic equations, and
    find the profile of the initial instability in the magnetization
    in the direction normal to the plane of the film. In general,
    this instability occurs at a finite wave number
    $k$. We determine the dependence of $k$ on the film thickness.
\end{abstract}
\maketitle
\section{Introduction}

    Experimental and theoretical studies of thin ferromagnetic films
    with an anisotropy axis perpendicular to the plane of the film
    have shown a diverse set of magnetic phases. In addition to
    uniformly magnetized states, strongly modulated (striped)
    patterns have been observed\cite{Davies04, Steren06, Donzelli03,Iunin06}
    as the dipolar stray field overcomes the exchange interaction
    and forces an oscillation of the magnetization. The anisotropy
    value at which the modulation first occurs for is known as the
    reorientation phase transition (RPT) point.
    \cite{Berger97,Kashuba93,Abanov95}

    This spontaneous modulation was first modeled by Garel and
    Doniach,\cite{Garel82} using the assumption that the local
    magnetization lies fully out of the plane of the film. The
    magnetization alternates between regions of uniform up
    and down polarization with width determined by the competition between the
    ferromagnetic exchange and the dipolar interactions. Refinements
    of this model have shown that near the RPT the domain structure
    is a cosine-like modulation of a nearly in-plane
    state,\cite{Kashuba93,Yafet88} rather than the sharply defined striped
    domains that occur at higher values of the out-of-plane
    anisotropy.

    Recently,\cite{Clarke07} we have found a universal phase diagram
    for zero-temperature transitions in thin film ferromagnets with
    perpendicular anisotropy and an external field applied
    perpendicular to the plane of the sample. In particular, we
    found analytic solutions for the boundary of metastability of
    uniformly magnetized phases in thin films. The uniformly up and
    down states decay along the lines $h=\mp\kappa$ respectively,
    while the canted phase decays at the boundary $(h/\kappa)^2 = 1+
    \kappa/\kappa_0$. These results, however, depend on the
    smallness of the parameter $\kappa_0 = t^2/(4\lambda)^2$, where
    $\lambda=\sqrt(2A/\mu_0M^2)$ is the exchange length. This
    approximation is not valid for all thin films being studied
    in current experiments. In particular, it fails in experiments
    with thermally evaporated thin nickel films,\cite{Lee06} where
    $\lambda\approx 20nm$ and $t$ ranges from $30nm$ to $200nm$.

    Three important effects come into play as the film thickness
    increases. First, the bound charges created on the top and
    bottom surfaces of the film by out-of-plane magnetization grow
    further apart, altering the strength of the dipolar interaction.
    Second, the magnetization may vary across the thickness of the
    film. Third, the direction of the in-plane magnetization may
    vary. The combination of the latter two effects allows the
    creation of closure domains that decrease the dipolar
    energy by following the local stray field. We find that at large
    thicknesses, the lowest energy modes are actually dominated by
    such domains.

    Variational studies of the dependence of the domain structure
    on the film thickness have been made,\cite{Marty00,Sukstanskii97}
    but have not allowed for variation of the magnetization along
    the direction normal to the film. While these studies show the
    correct dependence of the RPT anisotropy on thickness, they
    predict domain sizes that decrease monotonically with the
    thickness, opposite the actual trend at
    large thicknesses.\cite{O'Handley}

    This report examines the decay of uniform states in a thicker
    film via a spin wave analysis. We reduce the problem of finding
    the decay boundary of an in-plane state to a set of
    transcendental algebraic equations, and find the form of the
    initial decay modes. Section \ref{model} reviews the model for
    the magnetic free energy in a ferromagnetic film. Section
    \ref{waves} begins the spin wave analysis of the system around a
    uniformly polarized state, making general arguments about the
    shape and direction of the lowest energy fluctuations. In
    Section \ref{modes}, we find the governing eigenvalue equation
    for the spin wave modes in the $z$ direction (normal to the
    film) and find the general form of solutions to this equation.
    Section \ref{decay} completes the analysis by finding the mode
    with minimum energy and determines a set of equations that
    describe the reorientation phase transition point as a function
    of the thickness $t$.

\section{Model free energy}
\label{model}

    The free energy for a thin ferromagnetic film
    of thickness $t$ can be separated into local and long-range
    parts.  The local part includes the exchange, uniaxial
    anisotropy, and Zeeman terms: \begin{equation} E_\mathrm{local}=
    \int_{z=-t/2}^{t/2}\mathrm{d}^3r \, (A|\nabla \hat \mathbf m|^2
    - Km_z^2-\mu_0 M \, \mathbf H \cdot \hat \mathbf m),
    \label{eq-local} \end{equation} where $\hat \mathbf m =
    (\sin{\theta} \cos{\phi}, \, \sin{\theta} \sin{\phi}, \,
    \cos{\theta})$ is the 3-dimensional unit vector pointing along
    the magnetization. For simplicity, we shall
    consider the case $\mathbf{H}=0$.

    Experimentally, changing the temperature can serve to vary the
    anisotropy. It has been shown that the effective out-of-plane
    anisotropy decreases monotonically with increased
    temperature.\cite{Schulte95,Berger97, Kashuba93}

    The long-range part of the energy is due to dipolar interactions:
    \begin{equation}
        E_\mathrm{dip}=\frac{\mu_0M\!}{8\pi}^2\!
            \int\!\mathrm{d}^3r\!\int\!\mathrm{d}^3r'
            \,\frac{\rho(r)\rho(r')}{|\mathbf{r-r'}|},
        \label{eq-dipolar}
    \end{equation}
    where $\rho(r)=-\nabla\cdot(P(z)\mathbf{m}(\mathbf r))$.
    The integration runs over all space. Here $P(z)$ is the
    profile of the film in the $z$ direction. $P(z)=1$ if
    $-t/2\leq z\leq t/2$ and $0$ otherwise.

\section{Spin wave expansion around an in-plane state}
    \label{waves}

    Consider the free energy of small fluctuations around an
    in-plane state in which the magnetization points along the $y$
    axis:
    $\theta = \pi/2 +\nu(\mathbf{r})$,
    $\phi=\pi/2+\alpha(\mathbf{r})$,
    with $\nu$, $\alpha$ small.
    By transforming to Fourier space in the $xy$-plane, we can write
    the energy as an integral over modes that do not interact at
    second order in the fluctuations:
    \begin{equation}
        \Delta E=\int\frac{\mathrm{d}^2k}{(2\pi)^2}\left(
            \Delta E_{1}[\nu_\mathbf{k}]
            +\Delta E_{2}[\alpha_\mathbf{k}]
            +\Delta
            E_{\mathrm{dip}}[\nu_\mathbf{k},\alpha_\mathbf{k}]\right).
    \end{equation}
    This separation of modes is important because it means that the
    instability at the RPT will occur for a cosine-like fluctuation
    with a single wavelength. Here
    \begin{eqnarray}
        \frac{\Delta E_{1,\mathbf{k}}}{\mu_0M^2} &=&
            \int_{-t/2}^{t/2} \mathrm{d}z\left[
            \frac{\lambda^2}{2}\left|\nu'_{\mathbf{k}}\right|^2
            +\left|\nu_{\mathbf{k}}\right|^2\left(\frac{\lambda^2
            k^2}{2}-\frac{\kappa}{2}\right)
            \right],\nonumber\\
        \frac{\Delta E_{2,\mathbf{k}}}{\mu_0M^2} &=&
            \int_{-t/2}^{t/2} \mathrm{d}z\left[
            \frac{\lambda^2}{2}\left|\alpha'_{\mathbf{k}}\right|^2
            +\left|\alpha_{\mathbf{k}}\right|^2\frac{\lambda^2
            k^2}{2}\right],
    \end{eqnarray}
    and
    \begin{equation}
        \frac{\Delta
        E_{\mathrm{dip},\mathbf{k}}}{\mu_0M^2}=
            \frac{|k|}{4}\!\int_{-t/2}^{t/2}\!\!\!\!\!\mathrm{d}z\!\int_{-t/2}^{t/2}\!\!\!\!\!\mathrm{d}z'
            ~e^{-|k||z-z'|}f_{\mathbf{k}}(z,z')f_{-\mathbf{k}}(z',z),
    \end{equation}
    where
    \begin{equation}
    f_{\mathbf{k}}(z,z')=\mathrm{sgn}(z-z')\nu_{\mathbf{k}}(z)-i\frac{k_x}{|k|}\alpha_{\mathbf{k}}(z).
    \end{equation}
    Primes on $\nu$ and $\alpha$ denote $z$-derivatives. We define the effective
    anisotropy $\kappa=(K-\mu_0 M^2)/(\mu_0M^2/2)$ as the natural
    anisotropy less the uniform part of the dipolar interaction.

    Note that the dipolar energy couples symmetric fluctuations in
    $\nu$ to antisymmetric $\alpha$ fluctuations and vice-versa. In
    general, fluctuations in the azimuthal angle $\phi$ allow a gain
    in energy by compensating for the magnetic charges created by
    fluctuations in the polar angle $\theta$.

    It is these azimuthal fluctuations that cause the lowest energy
    mode to have its wave vector perpendicular to the initial
    magnetization (i.e. in the $x$-direction).  To see this,
    consider a mode such that the wave vector of the oscillation
    is at some angle $\beta$ to the $x$-axis, with a profile
    \mbox{$\{\alpha^{(0)}_{\mathbf{k}},\nu^{(0)}_{\mathbf{k}}\}$}
    for the azimuthal and polar fluctuations. For any such mode,
    we can find a corresponding fluctuation that has exactly
    the same dipolar energy but has its wave vector along the
    $x$-axis. Namely, \mbox{$\{\alpha^{(1)}_{\mathbf{k}},
    \nu^{(1)}_{\mathbf{k}}\}=\{\alpha^{(0)}_{\mathbf{k}}\cos\beta,\nu^{(0)}_{\mathbf{k}}\}$}.

    Further, since the azimuthal part of this new fluctuation
    is smaller by a factor of $\cos\beta$, it has a lower cost in
    the exchange energy. As a result, the difference in energy between
    the two types of fluctuations,
    \begin{equation}
        E^{(0)}_{\mathbf{k}}-E^{(1)}_{\mathbf{k}}=\sin^2\beta_0\int_{-t/2}^{t/2} \mathrm{d}z\left[
            \frac{\lambda^2}{2}\left|\alpha'^{(0)}_{\mathbf{k}}\right|^2
            +\left|\alpha^{(0)}_{\mathbf{k}}\right|^2\frac{\lambda^2
            k^2}{2}\right]
    \end{equation}
    is a nonnegative quantity.

    Since this is true of any profile for the azimuthal and polar
    fluctuations, it must be that the mode with the lowest energy
    cost has its wave vector oriented along the $x$-axis,
    perpendicular to the original direction of the magnetization.

    The exception to the above is when the wave vector of the
    oscillation lies along the $y$-axis, since the azimuthal
    fluctuation is then zero. In this case, however, no compensation
    can be made for the charges created by the fluctuations in
    $\theta$, so the energy will be inherently higher.

\section{Vertical spin wave eigenmodes}
\label{modes}
    Given that the lowest energy mode occurs with a wave vector along the $x$-axis, we work
    only in the subspace $k_y=0$.
    The energy in this subspace is diagonalized by modes that obey the eigenvalue (Lagrange) equations:
    \begin{eqnarray}
        \gamma\alpha_{\mathbf{k}}&=&-\lambda^2\alpha''_{\mathbf{k}}+k^2\lambda^2\alpha_{\mathbf{k}}\nonumber\\
            &+&\frac{k}{2}\int_{-t/2}^{t/2}\mathrm{d}z'
            ~e^{-|k||z-z'|}~f_{-\mathbf{k}}(z',z)\nonumber
    \end{eqnarray}
    and
    \begin{eqnarray}
        \gamma\nu_{\mathbf{k}}&=&-\lambda^2\nu''_{\mathbf{k}}+(k^2\lambda^2-\kappa)\nu_{\mathbf{k}}\nonumber\\
        &+&\frac{|k|}{2}\int_{-t/2}^{t/2}\mathrm{d}z'~e^{-|k||z-z'|}i\mathrm{sgn}(z'-z)
        f_{-\mathbf{k}}(z',z)\nonumber\\
            \label{mode-eq}
    \end{eqnarray}
    with $\alpha'(-t/2)=\alpha'(t/2)=\nu'(-t/2)=\nu'(t/2)=0$ induced by the
    boundary terms used in deriving the above equations.
    The energy of each mode is proportional to its eigenvalue
    $\gamma$. For a given value of $t$, the reorientation phase
    transition will occur at the $\kappa$ value when the lowest
    eigenvalue crosses zero, and some mode becomes soft. That is,
    at $\kappa$ such that $\gamma=\rm{d}\gamma/\rm{d}k=0$ for some mode.

    Note that the integral kernel in Eq.~(\ref{mode-eq}) is the
    solution to the differential equation
    $G''(z)-k^2G(z)=\delta(z)$, which has constant coefficients.
    Since (\ref{mode-eq}) is linear, and all the other terms in
    (\ref{mode-eq}) have constant coefficients as well,
     we would expect Eq.~(\ref{mode-eq}) to be solved by
    \begin{equation}
        \alpha_{\mathbf{k}}=\sum \alpha_{i,\mathbf{k}}e^{q_i
        z},\quad
        \nu_{\mathbf{k}}=\sum \nu_{i,\mathbf{k}}e^{q_i z},
        \label{trial-sol}
    \end{equation}
    where $\{\nu_{i,\mathbf{k}}\}$, $\{\alpha_{i,\mathbf{k}}\}$, and $q_i(k)$ are
    constants.

    This ansatz works so long as the $\{q_i\}$ are the solutions to the equation
    \begin{equation}
        a(a-\gamma)(a-\gamma-\kappa)=(\kappa+1)k^2\lambda^2,
        \label{ind-eq}
    \end{equation}
    where $a=(-q^2+k^2)\lambda^2$, and the
    conditions
    \begin{eqnarray}
        0&=&-iq_ik\nu_{i,\mathbf{k}}-(a_i^2-\gamma a_i+k^2)\alpha_{i,\mathbf{k}},\nonumber\\
        0 &=&\sum \frac{e^{q_it/2}}{|k|-q_i}
        \left(i|k|\nu_{i,\mathbf{k}}+k\alpha_{i,\mathbf{k}}\right),\nonumber\\
        0 &=&\sum \frac{e^{-q_it/2}}{|k|+q_i}
        \left(i|k|\nu_{i,\mathbf{k}}-k\alpha_{i,\mathbf{k}}\right).
        \label{constraint}
    \end{eqnarray}
    are satisfied.
    Note that there are six (not
    necessarily distinct) values for $q$, separated into three
    pairs $q=\pm\sqrt{k^2-a/\lambda^2}$.

    The boundary conditions on the top and bottom surfaces lead to
    \begin{eqnarray}
        0&=&\sum q_i e^{-q_it/2}\alpha_{i,\mathbf{k}}
         =\sum q_i e^{q_it/2}\alpha_{i,\mathbf{k}},\nonumber\\
        0&=&\sum q_i e^{-q_it/2}\nu_{i,\mathbf{k}}
         =\sum q_i e^{q_it/2}\nu_{i,\mathbf{k}}.
         \label{bound}
    \end{eqnarray}
    These boundary conditions, combined with Eqs.~(\ref{mode-eq})
    and (\ref{ind-eq}) are symmetric in the $z$-direction, allowing
    the separation of $\nu$ and $\alpha$ into modes with definite
    parity in the $z$-direction. A symmetric $\nu$ couples only to an
    antisymmetric $\alpha$ and vice-versa.
\section{Lowest Energy Eigenmode and the RPT}
\label{decay}
    In the thinnest films, $\nu$ is uniform across the thickness,
    and $\alpha$ is zero. As the thickness increases, we expect the
    parity of the lowest energy eigenmode to remain the same.
    The reason that a symmetric $\nu$ is favored is that the charges created
    on the top and bottom surfaces will be of opposite sign. This configuration has a
    lower energy cost than one with the same charges on the top and
    bottom surfaces, as would be the case for an antisymmetric
    $\nu$.

    If $\nu$ remains symmetric, then
    \begin{equation}
        \alpha_{\mathbf{k}}=\sum \tilde{\alpha}_{i,\mathbf{k}}\sinh(q_i
        z),\quad
        \nu_{\mathbf{k}}=\sum \tilde{\nu}_{i,\mathbf{k}}\cosh(q_i
        z),
    \end{equation}
    where the sum is over the three possible $a_i$, the positive
    branch is taken for $q_i$, and $\tilde{\alpha}_{i,\mathbf{k}}$ and $\tilde{\nu}_{i,\mathbf{k}}$
    are the appropriately symmetrized combinations of
    $\alpha_{i,\mathbf{k}}$ and $\nu_{i,\mathbf{k}}$.

    In this case, the constraint equations (\ref{constraint}) and (\ref{bound})
    reduce to the set
    \begin{eqnarray}
        0&\!\!=&\!\!\sum\left(\frac{|k|}{q_i}(a_i\!-\!\gamma+1)\cosh\frac{q_it}{2}
        +(a_i\!-\!\gamma)\sinh\frac{q_it}{2}\right)\tilde{\alpha}_{i,\mathbf{k}},\nonumber\\
        0&\!\!=&\!\!\sum \left(q_i\cosh\frac{q_it}{2}\right)\tilde{\alpha}_{i,\mathbf{k}},\nonumber\\
        0&\!\!=&\!\!\sum \left((a_i^2-\gamma
        a_i+k^2\lambda^2)\sinh\frac{q_it}{2}\right)\tilde{\alpha}_{i,\mathbf{k}}.
    \end{eqnarray}
    \begin{figure}[htbp]
        \begin{center}
            \includegraphics[width=0.9\columnwidth]{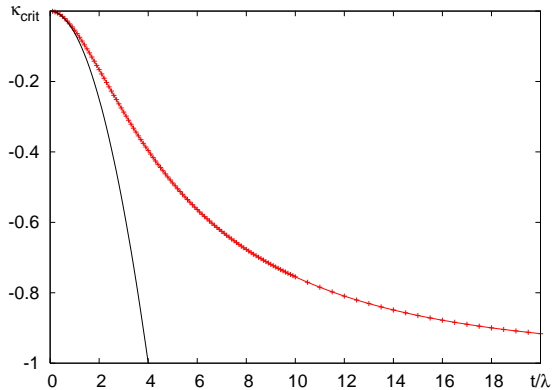}
        \end{center}
            \caption{Dependence of the reorientation phase
            transition anisotropy $\kappa_{\mathrm{crit}}$ on the film
            thickness. The line with $+$ marks is the exact
            solution, while the solid line is the thin film limit
            $\kappa_{\mathrm{crit}}=-(t/4\lambda)^2$}
        \label{fig-kap}
    \end{figure}
    In order to find the reorientation phase transition boundary, we
    need not solve for $\tilde{\alpha}_{i,\mathbf{k}}$, but we may
    rather note that in order for the above system to have a
    non-trivial solution, the determinant of the constraint
    conditions must be zero. Combining this with the condition
    $\gamma=\rm{d}\gamma/\rm{d}k=0$ for the RPT, we find that
    \begin{eqnarray}
        0&=&\epsilon_{lmn}D_{lmn},\nonumber\\
        0&=&\epsilon_{lmn}\frac{\mathrm{d}}{\mathrm{d}k}D_{lmn},
    \end{eqnarray}
    where
    \begin{eqnarray}
    D_{lmn}&=&\left(\frac{|k|}{q_l}(a_l+1)\cosh\frac{q_l t}{2}+a_l\sinh\frac{q_l t}{2}\right)\nonumber\\
        &\times&\!\!\left(q_m\cosh\frac{q_mt}{2}\right)\left((a_n^2+k^2\lambda^2)\sinh\frac{q_nt}{2}\right).
    \end{eqnarray}
    \begin{figure}[htbp]
        \begin{center}
            \includegraphics[width=0.9\columnwidth]{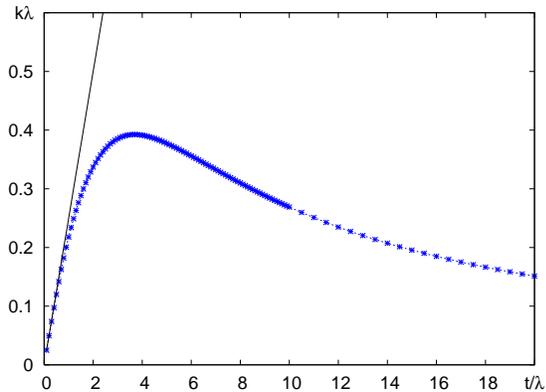}
        \end{center}
            \caption{Dependence of the instability wavenumber $k$
            on thickness. The starred line is the exact solution.
            The solid line is the thin film limit
            $k\rightarrow t/4\lambda^2$.  }
        \label{fig-k}
    \end{figure}
    Together with Eq.~(\ref{ind-eq}) for the $q$
    values, this completes the system of equations necessary to find
    both the $\kappa_{\mathrm{crit}}$ value of the RPT and the initial
    wavenumber $k$ of the initial oscillation. Since these equations are
    transcendental, no further analytic progress may be made. A
    plot of the numerically determined solution to these equations
    is found in Figs.~\ref{fig-kap} and \ref{fig-k}, along with
    a comparison to the previously determined thin-film
    limit\cite{Clarke07}. The thin film limit
    $\kappa_{\mathrm{crit}}\rightarrow-(t/4\lambda)^2$, $k\rightarrow t/4\lambda^2$ can be seen to
    work well for thicknesses $t \lesssim 2\lambda$. As the
    thickness increases, the initial wavenumber reaches a maximum
    (when $t/\lambda=3.7\pm.1$), and decreases to zero (uniform
    rotation) for infinite thickness. For large thicknesses, the wavenumber
    goes like $\pi/t$, which is consistent with the
    linear dependence of domain size on thickness in thicker films.\cite{O'Handley}
     The critical value of the effective
    anisotropy changes monotonically from its thin film limit to the
    bulk limit $\kappa=-1$ at $t=\infty$.

    The dependence of the entire canted phase boundary on
    thickness may be derived in a similar fashion. Further
    complications arise, however, as the characteristic equation
    (\ref{ind-eq}) loses the symmetry that allows us to solve it as
    a cubic, and becomes fully sixth order. Further, when the initial
    magnetization is tilted, the mode with the lowest eigenvalue
    will no longer have its wavevector parallel to the $x$-axis.

    \begin{figure}[htbp]
        \includegraphics[width=.1\columnwidth, height=.1\columnwidth]{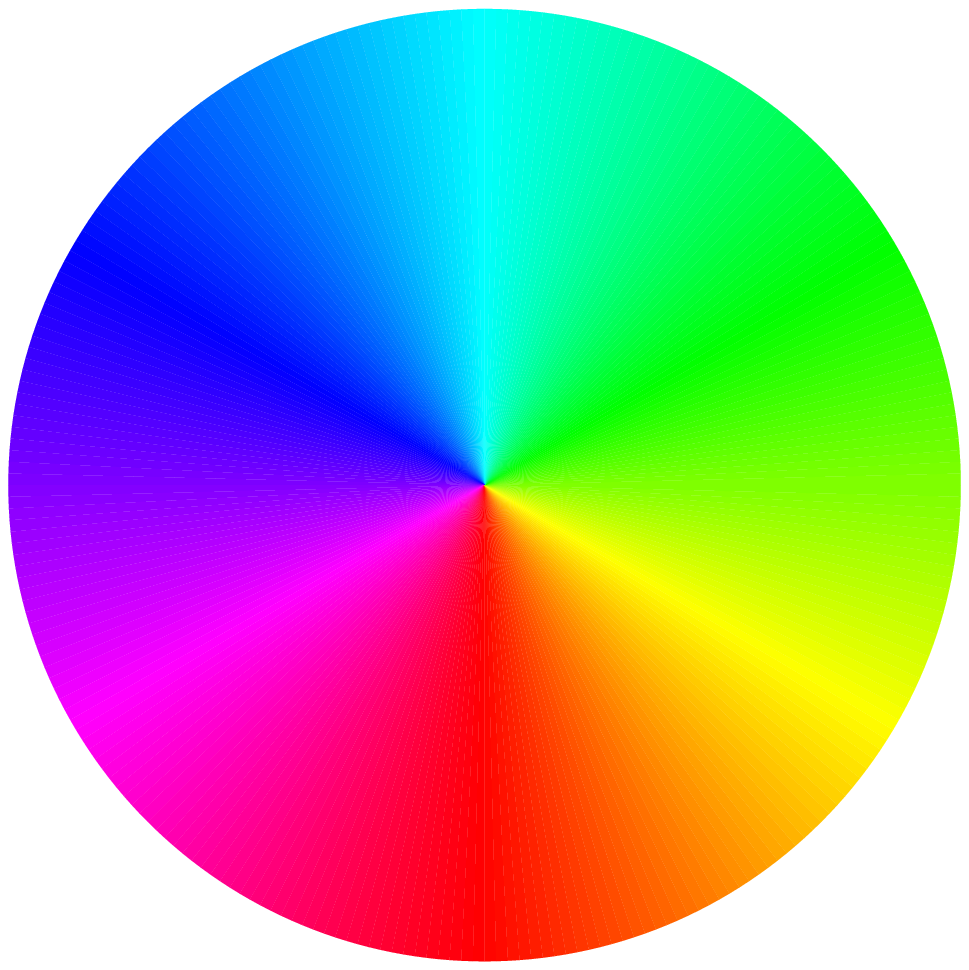}
        \begin{center}
            $t=.5\lambda$
            \includegraphics[width=\columnwidth, height=.01667\columnwidth]{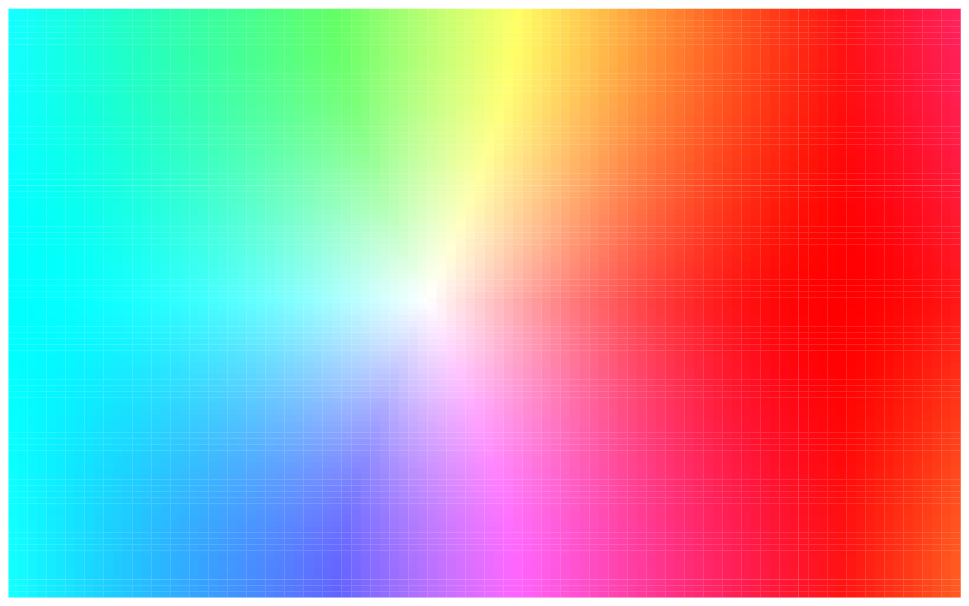}
            $t=\lambda$
            \includegraphics[width=\columnwidth, height=.03333\columnwidth]{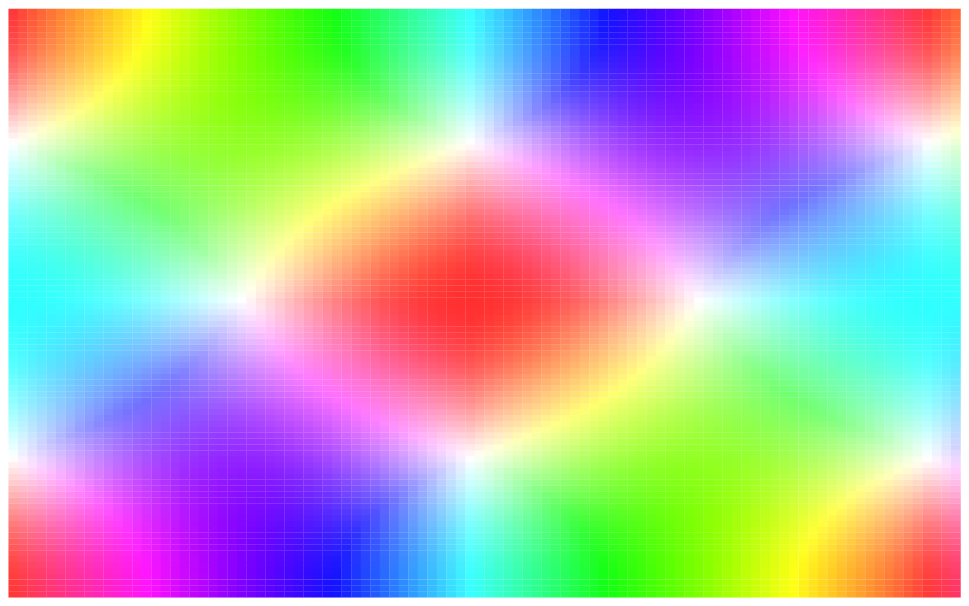}
            $t=1.5\lambda$
            \includegraphics[width=\columnwidth, height=.05\columnwidth]{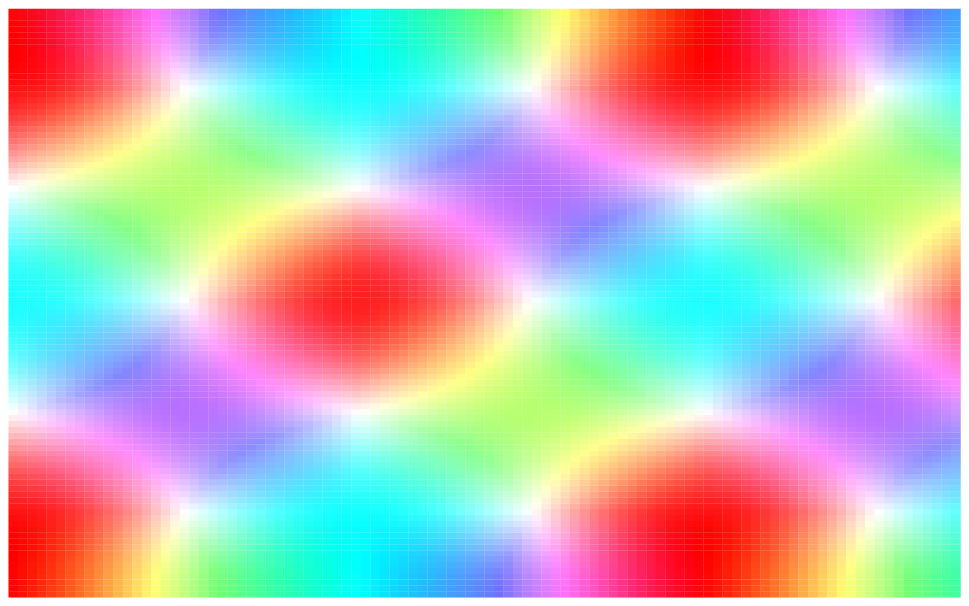}
            $t=2\lambda$
            \includegraphics[width=\columnwidth, height=.06667\columnwidth]{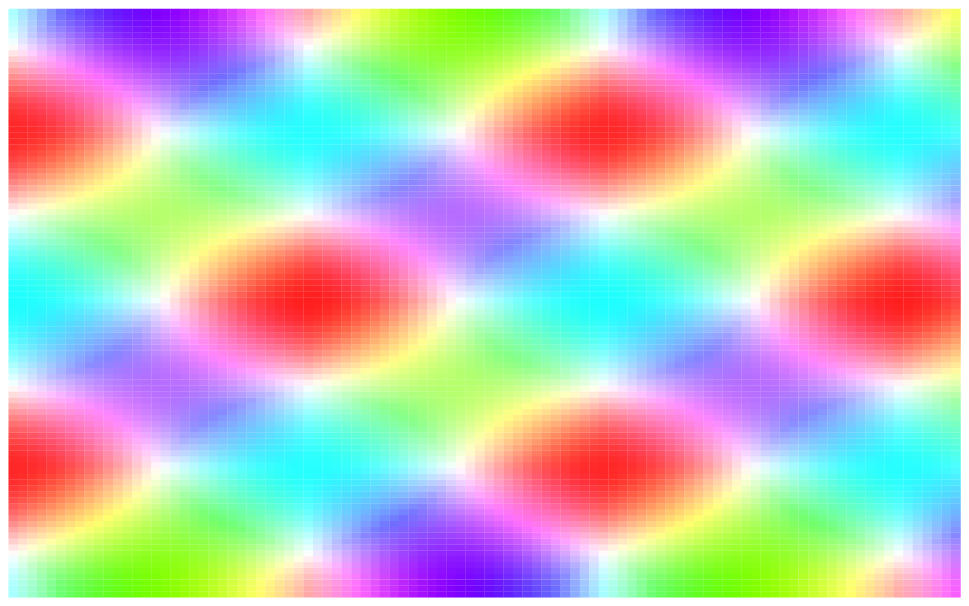}
            $t=5\lambda$
            \includegraphics[width=\columnwidth, height=.16667\columnwidth]{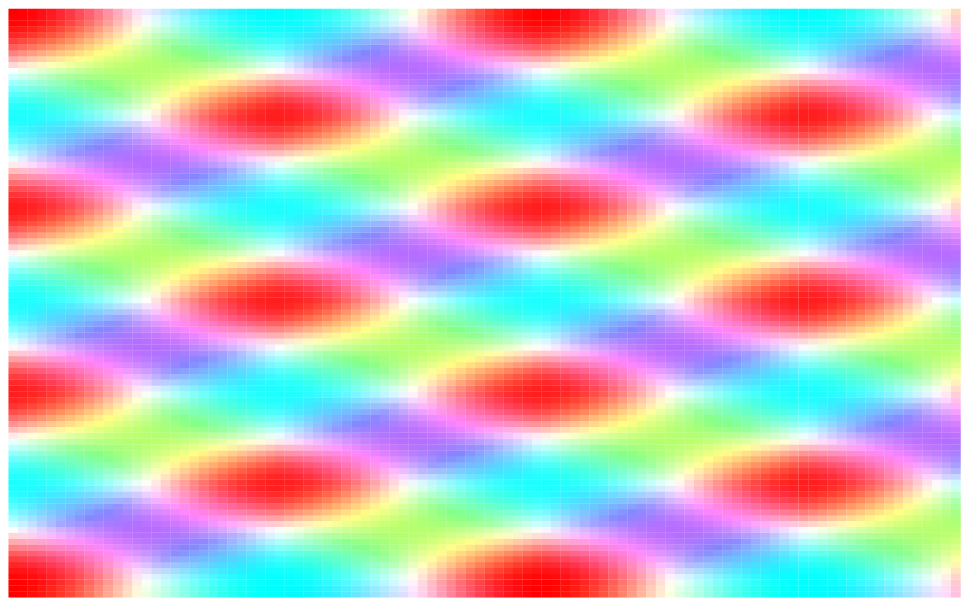}
        \end{center}
        \caption{First unstable modes for various thicknesses.
            The film is shown in profile, such that
            magnetization is mostly into the page, with small
            fluctuations in the directions indicated by the color
            wheel. The modes are periodic in the $x$-direction, which
            runs across the page. A section $30\lambda$ wide is
            shown.}
        \label{domain-fig}
    \end{figure}
    \section{Conclusions}
    In this report, we have developed the thickness dependence of
    the reorientation phase transition, deriving values for
    $\kappa_{\mathrm{crit}}(t)$ and $k(t)$ that may be directly compared with
    experiment. In so doing, we have connected recently derived thin
    film limits with the well-known linear dependence of domain size on
    thickness in thicker films.

        Fig.~\ref{domain-fig} shows the fluctuation patterns in the
    modes that first become unstable for a variety of thicknesses.
    In the thinnest films, the assumption that the magnetization is
    uniform across the thickness of the film is valid. The azimuthal
    angle is nearly zero everywhere. As the thickness
    increases, however, this assumption breaks down, leading to a drastically
    different landscape of magnetization as closure domains form
    near the upper and lower surfaces of the film. For very thick
    films, the lowest energy mode can have many repetitions of this
    closure behavior.

    As the anisotropy increases well beyond the RPT point, it is likely
    that many of these repetitions would be forced out, to
    accommodate the demand for more vertical magnetization. However,
    at the RPT point itself, this leads for very thick films to an
    essentially periodic structure in the $z$-direction.

    Since the transition away from the in-plane state as the
    anisotropy increases is continuous,\cite{Kashuba93,Clarke07}
    the ground state of a magnetic film that is just beyond the
    transition point should look much like the modes in
    Fig.~\ref{domain-fig} that first become unstable. For such films
    we predict a strong, nearly oscillatory thickness dependence of the
    stray magnetic field, as the surface of the film becomes
    dominated alternately by out-of-plane magnetization, as for
    $t=1.5\lambda$ and $t=5\lambda$, and in-plane closure domains,
    as for $t=\lambda$ and $t=2\lambda$. This effect may
    be observable with magnetic force microscopy.

\section{Acknowledgements}
    Thanks to M. McEvoy, O. A. Tretiakov and O. Tchernyshyov for helpful
    discussions, and to S.H. Lee, F. Q. Zhu, C.~L. Chien, and N. Markovic for
    sharing their unpublished data. This work was supported in part by NSF Grant DMR-0520491.


\begin{thebibliography}{32}
\expandafter\ifx\csname
natexlab\endcsname\relax\def\natexlab#1{#1}\fi
\expandafter\ifx\csname bibnamefont\endcsname\relax
  \def\bibnamefont#1{#1}\fi
\expandafter\ifx\csname bibfnamefont\endcsname\relax
  \def\bibfnamefont#1{#1}\fi
\expandafter\ifx\csname citenamefont\endcsname\relax
  \def\citenamefont#1{#1}\fi
\expandafter\ifx\csname url\endcsname\relax
  \def\url#1{\texttt{#1}}\fi
\expandafter\ifx\csname urlprefix\endcsname\relax\def\urlprefix{URL
}\fi \providecommand{\bibinfo}[2]{#2}
\providecommand{\eprint}[2][]{\url{#2}}

\bibitem[{\citenamefont{J. E. Davies et~al.}(2004)\citenamefont{Davies, Hellwig, Fullerton,
Denbeaux, Kortright and Liu}}]{Davies04}
\bibinfo{author}{\bibfnamefont{J.~E.}~\bibnamefont{Davies}},
  \bibinfo{author}{\bibfnamefont{O.}~\bibnamefont{Hellwig}},
  \bibinfo{author}{\bibfnamefont{E.~E}~\bibnamefont{Fullerton}},
  \bibinfo{author}{\bibfnamefont{G.}~\bibnamefont{Denbeaux}},
  \bibinfo{author}{\bibfnamefont{J.~B.}~\bibnamefont{Kortright}},
  \bibnamefont{and} \bibinfo{author}{\bibfnamefont{K.}~\bibnamefont{Liu}},
  \bibinfo{journal}{Phys. Rev. B} \textbf{\bibinfo{volume}{70}},
  \bibinfo{pages}{224434} (\bibinfo{year}{2004}).

\bibitem[{\citenamefont{L. B. Steren et~al.}(2006)\citenamefont{Steren, Milano, Garcia,
Marangolo, Eddrief, and Etgens}}]{Steren06}
\bibinfo{author}{\bibfnamefont{L.~B.}~\bibnamefont{Steren}},
  \bibinfo{author}{\bibfnamefont{J.}~\bibnamefont{Milano}},
  \bibinfo{author}{\bibfnamefont{V.}~\bibnamefont{Garcia}},
  \bibinfo{author}{\bibfnamefont{M.}~\bibnamefont{Marangolo}},
  \bibinfo{author}{\bibfnamefont{M.}~\bibnamefont{Eddrief}},
  \bibnamefont{and} \bibinfo{author}{\bibfnamefont{V.~H.}~\bibnamefont{Etgens}},
  \bibinfo{journal}{Phys. Rev. B} \textbf{\bibinfo{volume}{74}},
  \bibinfo{pages}{144402} (\bibinfo{year}{2006}).

\bibitem[{\citenamefont{O. Donzelli et~al.}(2006)\citenamefont{Donzelli, Palmeri, Musa,
Casoli, Albertini, Pareti, and Turilli}}]{Donzelli03}
\bibinfo{author}{\bibfnamefont{O.}~\bibnamefont{Donzelli}},
\bibinfo{author}{\bibfnamefont{D.}~\bibnamefont{Palmeri}},
  \bibinfo{author}{\bibfnamefont{L.}~\bibnamefont{Musa}},
  \bibinfo{author}{\bibfnamefont{F.}~\bibnamefont{Casoli}},
  \bibinfo{author}{\bibfnamefont{F.}~\bibnamefont{Albertini}},
  \bibinfo{author}{\bibfnamefont{L.}~\bibnamefont{Pareti}},
  \bibnamefont{and} \bibinfo{author}{\bibfnamefont{G.}~\bibnamefont{Turilli}},
  \bibinfo{journal}{J. Appl. Phys.} \textbf{\bibinfo{volume}{93}},
  \bibinfo{pages}{9908} (\bibinfo{year}{2003}).

\bibitem{Iunin06}
Y. L. Iunin, Y. P. Kabanov, V. I. Nikitenko, X. M. Cheng, D. Clarke,
O.~A. Tretiakov, O. Tchernyshyov, A.~J. Sharpiro, R.~D. Shull, and
C.~ L. Chien, Phys. Rev. Lett. 98, 117204 (2007).

\bibitem[{\citenamefont{A. Berger and R. P. Erickson}(1997)\citenamefont{Berger and Erickson}}]{Berger97}
\bibinfo{author}{\bibfnamefont{A.}~\bibnamefont{Berger}}
  \bibnamefont{and} \bibinfo{author}{\bibfnamefont{R.~P.}~\bibnamefont{Erickson}},
  \bibinfo{journal}{J. Magn. Magn. Mater.} \textbf{\bibinfo{volume}{165}},
  \bibinfo{pages}{70} (\bibinfo{year}{1997}).
\bibitem[{\citenamefont{A. B. Kashuba and V. L. Pokrovsky}(1993)\citenamefont{Kashuba and Pokrovsky}}]{Kashuba93}
\bibinfo{author}{\bibfnamefont{A.~B.}~\bibnamefont{Kashuba}}
  \bibnamefont{and} \bibinfo{author}{\bibfnamefont{V.~L.}~\bibnamefont{Pokrovsky}},
  \bibinfo{journal}{Phys. Rev. B} \textbf{\bibinfo{volume}{48}},
  \bibinfo{pages}{10335} (\bibinfo{year}{1993}).
\bibitem[{\citenamefont{Ar. Abanov et~al.}(1995)\citenamefont{Ar. Abanov, V. Kalatsky, V. L. Pokrovsky, and  Saslow}}]{Abanov95}
\bibinfo{author}{\bibfnamefont{Ar.}~\bibnamefont{Abanov}},
  \bibinfo{author}{\bibfnamefont{V.}~\bibnamefont{Kalatsky}},
  \bibinfo{author}{\bibfnamefont{V.~L.}~\bibnamefont{Pokrovsky}},
  \bibnamefont{and} \bibinfo{author}{\bibfnamefont{W.~M.}~\bibnamefont{Saslow}},
  \bibinfo{journal}{Phys. Rev. B} \textbf{\bibinfo{volume}{51}},
  \bibinfo{pages}{1023} (\bibinfo{year}{1995}).

\bibitem[{\citenamefont{T. Garel and S. Doniach}(1982)\citenamefont{Garel and Doniach}}]{Garel82}
\bibinfo{author}{\bibfnamefont{T.}~\bibnamefont{Garel}}
  \bibnamefont{and} \bibinfo{author}{\bibfnamefont{S.}~\bibnamefont{Doniach}},
  \bibinfo{journal}{Phys. Rev. B} \textbf{\bibinfo{volume}{26}},
  \bibinfo{pages}{325} (\bibinfo{year}{1982}).


\bibitem[{\citenamefont{Y. Yafet and E. M. Gyorgy}(1988)\citenamefont{Yafet and Gyorgy}}]{Yafet88}
\bibinfo{author}{\bibfnamefont{Y.}~\bibnamefont{ Yafet }}
  \bibnamefont{and} \bibinfo{author}{\bibfnamefont{E.~M.}~\bibnamefont{Gyorgy}},
  \bibinfo{journal}{Phys. Rev. B} \textbf{\bibinfo{volume}{38}},
  \bibinfo{pages}{9145} (\bibinfo{year}{1988}).


\bibitem[{\citenamefont{ D. Clarke, O. A. Tretiakov, and O. Tchernyshyov}(2007)
\citenamefont{Lee, Zhu, Chien, and Markovic}}]{Clarke07}
  \bibinfo{author}{\bibfnamefont{D.}~\bibnamefont{Clarke}},
  \bibinfo{author}{\bibfnamefont{O.~A.}~\bibnamefont{Tretiakov}},
  \bibnamefont{and}
  \bibinfo{author}{\bibfnamefont{O.}~\bibnamefont{Tchernyshyov}},Phys. Rev. B
  (To be published) cond-mat/0612346

\bibitem[{\citenamefont{ S.~H. Lee, F. Q. Zhu, C.~L. Chien, N. Markovic}(2006)
\citenamefont{Lee, Zhu, Chien, and Markovic}}]{Lee06}
  \bibinfo{author}{\bibfnamefont{S.~H.}~\bibnamefont{Lee}},
  \bibinfo{author}{\bibfnamefont{F.~Q.}~\bibnamefont{Zhu}},
  \bibinfo{author}{\bibfnamefont{C.~L.}~\bibnamefont{Chien}},
  \bibnamefont{and}
  \bibinfo{author}{\bibfnamefont{N.}~\bibnamefont{Markovic}},
  unpublished.


\bibitem[{\citenamefont{A. Marty et~al.}(2006)\citenamefont{ Marty, Samson, Gilles,
Belakhovsky, Dudzik, Durr, Dhesi, van der Laan, and
Goedkoop}}]{Marty00}
\bibinfo{author}{\bibfnamefont{A.}~\bibnamefont{Marty}},
\bibinfo{author}{\bibfnamefont{Y.}~\bibnamefont{Samson}},
  \bibinfo{author}{\bibfnamefont{B.}~\bibnamefont{Gilles}},
  \bibinfo{author}{\bibfnamefont{M.}~\bibnamefont{Belakhovsky}},
  \bibinfo{author}{\bibfnamefont{E.}~\bibnamefont{Dudzik}},
  \bibinfo{author}{\bibfnamefont{H.}~\bibnamefont{Durr}},
  \bibinfo{author}{\bibfnamefont{S.~S.}~\bibnamefont{Dhesi}},
  \bibinfo{author}{\bibfnamefont{G.}~\bibnamefont{van der Laan}},
  \bibnamefont{and} \bibinfo{author}{\bibfnamefont{J.~B.}~\bibnamefont{Goedkoop}},
  \bibinfo{journal}{J. Appl. Phys.} \textbf{\bibinfo{volume}{87}},
  \bibinfo{pages}{5472} (\bibinfo{year}{2000}).

\bibitem[{\citenamefont{Sukstanskii et~al.}(2006)\citenamefont{ Sukstanskii and
Primak}}]{Sukstanskii97}
\bibinfo{author}{\bibfnamefont{A.~L.}~\bibnamefont{Sukstanskii}},
  \bibnamefont{and} \bibinfo{author}{\bibfnamefont{K.~I.}~\bibnamefont{Primak}},
  \bibinfo{journal}{J. Magn. Magn. Mater} \textbf{\bibinfo{volume}{169}},
  \bibinfo{pages}{31} (\bibinfo{year}{1997}).


\bibitem[{\citenamefont{R.~C. O'Handley}(2000)
\citenamefont{R.~C. O'Handley}}]{O'Handley}
\bibinfo{author}{\bibfnamefont{R.~C.}~\bibnamefont{O'Handley}},
\emph{Modern Magnetic Materials: Principles and Applications} (Wiley
\& Sons, New York, 2000) p. 655.

\bibitem[{\citenamefont{O. Schulte et~al.}(1995)\citenamefont{Schulte, Klose, and Felsch}}]{Schulte95}
\bibinfo{author}{\bibfnamefont{O.}~\bibnamefont{Schulte}},
  \bibinfo{author}{\bibfnamefont{F.}~\bibnamefont{Klose}},
  \bibnamefont{and} \bibinfo{author}{\bibfnamefont{W.}~\bibnamefont{Felsch}},
  \bibinfo{journal}{Phys. Rev. B} \textbf{\bibinfo{volume}{52}},
  \bibinfo{pages}{6480} (\bibinfo{year}{1995}).







\end{thebibliography}
\end{document}